# Comparison of theoretical elastic couple stress predictions with physical experiments for pure torsion

Ali R. Hadjesfandiari, Gary F. Dargush

*Department of Mechanical and Aerospace Engineering*
*University at Buffalo, State University of New York*
*Buffalo, NY 14260 USA*

ah@buffalo.edu, gdargush@buffalo.edu

April 24, 2016

## Abstract

Several different versions of couple stress theory have appeared in the literature, including the indeterminate Mindlin-Tiersten-Koiter couple stress theory (MTK-CST), indeterminate symmetric modified couple stress theory (M-CST) and determinate skew-symmetric consistent couple stress theory (C-CST). First, the solutions within each of these theories for pure torsion of cylindrical bars composed of isotropic elastic material are presented and found to provide a remarkable basis for comparison with observed physical response. In particular, recent novel physical experiments to characterize torsion of micro-diameter copper wires in quasi-static tests show no significant size effect in the elastic range. This result agrees with the prediction of the skew-symmetric C-CST that there is no size effect for torsion of an elastic circular bar in quasi-static loading, because the mean curvature tensor vanishes in a pure twist deformation. On the other hand, solutions within the other two theories exhibit size-dependent torsional response, which depends upon either one or two additional material parameters, respectively, for the indeterminate symmetric M-CST or indeterminate MTK-CST. Results are presented to illustrate the magnitude of the expected size-dependence within these two theories in torsion. Interestingly, if the material length scales for copper in these two theories with size-dependent torsion is on the order of microns or larger, then the recent physical experiments in torsion would align only with the self-consistent skew-symmetric couple stress theory, which inherently shows no size effect.

## 1. Introduction

Recently, in References [1, 2], the authors have developed consistent size-dependent continuum mechanics theory for solids, which can be regarded as a natural extension of rigid body mechanics. In this theory, the rigid body portion of motions of infinitesimal elements (or rigid triads) at each point of the continuum are considered. More specifically, the underlying kinematics of deformation is based on the relative translation and rotation of infinitesimal elements of matter at adjacent points of the continuum, thus capturing the deformation in terms of stretches and curvatures, respectively. Another fundamental step in this development is satisfying the requirement that the normal component of the couple-traction vector must vanish on the boundary surface in a systematic way, which establishes the skew-symmetric character of the couple-stress tensor. Consequently, the stresses are fully determinate, and the measure of deformation energetically-conjugate to couple-stress is the skew-symmetrical mean curvature tensor, which describes bending deformation. This consistent couple stress theory (C-CST) involves only true continuum kinematical quantities without recourse to any additional artificial degrees of freedom. An initial incomplete version of couple stress theory was developed more than a half-century ago by Mindlin and Tiersten [3] and Koiter [4], which used the gradient of the rotation as the curvature tensor measure. However, this curvature tensor is not a proper measure of deformation and creates indeterminacy of the couple-stress tensor in this Mindlin-Tiersten-Koiter (MTK) theory. The symmetric modified couple stress theory (M-CST) of Yang et al. [5] still suffers from the same inconsistencies and difficulties with the underlying formulation. Furthermore, the original development of M-CST relies upon a fictitious moment of couples balance law, which places the whole theory on a most precarious foundation [6]. In addition, the measure of deformation energetically-conjugate to the symmetric couple-stress tensor is the symmetric torsional curvature tensor [1, 2, 7] that is unable to represent a bending deformation properly.

Although it is important for a continuum theory to be mathematically consistent, it has to be validated by careful experimentation or observation. Therefore, any proposed continuum theory must be compared critically with a number of detailed physical experiments designed to test predicted new features of the mechanical behavior of solids. In the present paper, we focus on this stage by comparing analytical torsion solutions from the different couple stress theories with available results from recent experiments on torsion of polycrystalline copper wires of micro-diameters. Historically, bending and torsion testing are the traditional experiments to study mechanical behavior of solids in static and quasi-static loading. Although all of the available size-dependent theories, such as the variety of couple stress theories, strain gradient theories and micropolar theories, predict very similar bending size effects on micro-size scales, the theories predict quite different behavior in torsion. Particularly, the skew-symmetric consistent couple stress theory (C-CST) predicts no size effect in the elastic range for pure torsion of a circular isotropic elastic bar in static and quasi-static loading [1]. Therefore, torsion testing is very important in studying the mechanical behavior of micro-size characteristic length samples to verify the consistency of different size-dependent continuum theories. Furthermore, it has already been recognized that torsion testing on thin metal wires is one of the most fundamental approaches to assessing the mechanical behavior of materials at small scales [8]. This covers the entire deformation range, from elastic deformation, through yielding, to the hardening regime.

Recent torsion experiments on micro-diameter wires in quasi-static loading can be considered as extensions of the seminal experiments of Fleck et al. [9], who developed a special screw driven system to conduct quasi-static torsion experiments on polycrystalline micro-diameter copper wires in the plastic strain range. The Fleck et al. [9] torsion experimental results showed significant size-effect in the plastic range. On the other hand, experiments by Lu and Song [10] and Song and Lu [11], who improved the Fleck et al. technique, showed no significant size effect in both the elastic and plastic range for torsion of micro-diameter polycrystalline copper wires under quasi-static loading. Liu et al. [8] also followed Fleck et al. [9] with some modification in the experimental technique to characterize torsion of micro-diameter polycrystalline copper wires in quasi-static loading and showed no significant size effect in the elastic range. However, their results showed strong size effect in both the initial yielding and the plastic flow.

In this paper, we review these physical experiments by Liu et al. [8], Lu and Song [10] and Song and Lu [11], which agree completely with the prediction of skew-symmetric consistent couple stress theory (C-CST) that there is no size effect in the elastic range for pure torsion of micro-diameter copper wires in quasi-static loading. Importantly, the predictions of indeterminate MTK-CST and indeterminate M-CST disagree qualitatively with these experiments. However, a word of caution is in order, because the results for all of these theories depend on elastic parameters that can be viewed in terms of characteristic material lengths, which are not well-established for poly-crystalline copper. If these length scales are sufficiently small, say on the order of 100 nm or less, then the size effect in torsion would not be observable in micro-diameter copper wires.

The remainder of the paper is organized as follows. In Sect. 2, we give a brief review of torsional deformation in general, and present the kinematics of pure torsion for a circular bar. In Sect. 3, we examine the pure torsional deformation of the circular bar in the linear isotropic classical and couple stress theories of elasticity, which include the original indeterminate MTK-CST, indeterminate symmetric M-CST and determinate skew-symmetric C-CST. Section 4 presents a review of previously published torsion experiments, which provide evidence consistent with the skew-symmetric couple stress theory. Finally, Sect. 5 contains a summary and some general conclusions.

## 2. Torsional deformation

Consider the $z$-axis of the coordinate system along the axis of a circular bar with constant cross section and isotropic material response. Let the radius and length of the bar be $a$ and $L$, respectively.

For a general torsional deformation in cylindrical coordinates $r,\theta,z$, only the rotational displacement component $u_\theta$ is non-zero. Therefore, we have

$$u_r = 0, \qquad u_\theta = u_\theta\left(r,z\right), \qquad u_z = 0 \tag{1a-c}$$

From this we obtain

$$e_{rr} = e_{\theta\theta} = e_{zz} = e_{rz} = 0 \tag{2}$$

and the non-zero components of strains are

$$e_{\theta z} = \frac{1}{2}\frac{\partial u_\theta}{\partial z}, \quad e_{r\theta} = \frac{1}{2}\left(\frac{\partial u_\theta}{\partial r} - \frac{u_\theta}{r}\right) \tag{3a,b}$$

We notice that the components of the rotation vector are

$$\omega_r = \omega_{z\theta} = -\frac{1}{2}\frac{\partial u_\theta}{\partial z} \tag{4a}$$

$$\omega_\theta = \omega_{rz} = 0 \tag{4b}$$

$$\omega_z = \omega_{\theta r} = \frac{1}{2}\left(\frac{\partial u_\theta}{\partial r} + \frac{u_\theta}{r}\right) \tag{4c}$$

The general form of the gradient of the rotation vector in cylindrical coordinates is given by

$$[\nabla\omega] = \begin{bmatrix} \frac{\partial \omega_r}{\partial r} & \frac{1}{r}\left(\frac{\partial \omega_r}{\partial \theta} - \omega_\theta\right) & \frac{\partial \omega_r}{\partial z} \\ \frac{\partial \omega_\theta}{\partial r} & \frac{1}{r}\left(\frac{\partial \omega_\theta}{\partial \theta} + \omega_r\right) & \frac{\partial \omega_\theta}{\partial z} \\ \frac{\partial \omega_z}{\partial r} & \frac{1}{r}\frac{\partial \omega_\theta}{\partial \theta} & \frac{\partial \omega_z}{\partial z} \end{bmatrix} \tag{5}$$

This tensor is important for defining the curvature tensor in different couple stress theories. For general torsional deformation, this tensor reduces to

$$[\nabla\omega] = \begin{bmatrix} \frac{\partial \omega_r}{\partial r} & 0 & \frac{\partial \omega_r}{\partial z} \\ 0 & \frac{\omega_r}{r} & 0 \\ \frac{\partial \omega_z}{\partial r} & 0 & \frac{\partial \omega_z}{\partial z} \end{bmatrix} \tag{6}$$

For the special case of pure torsion or twisting of a uniform bar, the rotational or peripheral displacement $u_\theta$ becomes

$$u_\theta(r,z) = \alpha r z \tag{7}$$

where $\alpha$ is the constant angle of twist per unit length. This constant is defined as

$$\alpha = \frac{\phi}{L} \tag{8}$$

where $\phi$ is the rotation of the end at $z = L$ relative to the fixed end at $z = 0$, called the twist angle. The non-zero components of the strain tensor and rotation vector for this deformation are

$$e_{\theta z} = \frac{1}{2}\alpha r \, , \quad e_{r\theta} = 0 \tag{9a,b}$$

$$\omega_r = \omega_{z\theta} = -\frac{1}{2}\alpha r \, , \qquad \omega_z = \omega_{\theta r} = \alpha z \tag{10a,b}$$

Therefore, the rotation gradient for pure torsion in cylindrical polar coordinates becomes

$$[\nabla \omega] = \begin{bmatrix} -\frac{1}{2}\alpha & 0 & 0 \\ 0 & -\frac{1}{2}\alpha & 0 \\ 0 & 0 & \alpha \end{bmatrix} \tag{11}$$

As we know, this deformation describes the twist of the cylindrical bar in small deformation classical linear isotropic elasticity.

## 3. Pure torsion in classical and couple stress elasticity theories

Now we examine the pure torsional deformation of the circular bar in the linear isotropic classical and couple stress theories of elasticity.

The force and moment balance equations for general couple stress theory under quasi-static conditions in the absence of body forces are

$$\sigma_{ji,j} = 0 \tag{12}$$

$$\mu_{ji,j} + \varepsilon_{ijk}\sigma_{jk} = 0 \tag{13}$$

Here $\sigma_{ij}$ represents the force-stress tensor, $\mu_{ij}$ is the couple-stress tensor and $\varepsilon_{ijk}$ is the Levi-Civita alternating symbol.

### *3.1. Cauchy (classical) elasticity*

In this theory, there are no couple-stresses and the force-stresses are symmetric, that is

$$\mu_{ij} = 0 \quad , \quad \sigma_{ji} = \sigma_{ij} \tag{14a,b}$$

As a result, the constitutive relation for linear isotropic elastic material is

$$\sigma_{ij} = \lambda e_{kk}\delta_{ij} + 2Ge_{ij} \tag{15}$$

where the moduli $\lambda$ and $G$ are the Lamé constants for isotropic media, and $G$ is also referred to as the shear modulus. These two constants are related by

$$\lambda = 2G\frac{\nu}{1-2\nu} \tag{16}$$

where $\nu$ is the Poisson ratio. The positive-definite elastic energy condition requires

$$3\lambda + 2G > 0, \qquad G > 0 \tag{17a,b}$$

Consequently, the final governing equations for linear isotropic elastic material in terms of displacements are

$$(\lambda + G)u_{k,ki} + G\nabla^2 u_i = 0 \tag{18}$$

For the pure torsion deformation, the non-zero components of force-stresses are

$$\sigma_{z\theta} = \sigma_{\theta z} = G\alpha r \tag{19}$$

We notice that these stresses satisfy the equilibrium equations and free stress surface boundary condition on the cylindrical surface of the bar. The resultant torque at the ends of the bar can be obtained as

$$T_\sigma = \int_0^a \sigma_{z\theta} 2\pi r^2 dr = \frac{1}{2}\pi G\alpha a^4 \tag{20}$$

### *3.2. Indeterminate Mindlin-Tiersten-Koiter couple stress theory (MTK-CST)*

In this theory, the curvature tensor is given by

$$k_{ij} = \omega_{j,i} \tag{21}$$

and the constitutive relations for linear isotropic elastic material are [3]

$$\sigma_{(ij)} = \lambda e_{kk}\delta_{ij} + 2Ge_{ij} \tag{22}$$

$$\begin{aligned} \mu_{ij} &= Q\delta_{ij} + 4\eta k_{ij} + 4\eta' k_{ji} \\ &= Q\delta_{ij} + 4\eta\omega_{j,i} + 4\eta'\omega_{i,j} \end{aligned} \tag{23}$$

Here $\sigma_{(ij)}$ is the symmetric part of force-stress tensor and $Q\delta_{ij}$ is the indeterminate spherical part of the couple-stress tensor, where $Q$ is a pseudo-scalar. The constants $\eta$ and $\eta'$ are the couple stress material constants for isotropic media. We can define the ratios

$$\frac{\eta}{G} = l^2, \qquad \frac{\eta'}{\eta} = c \tag{24a,b}$$

Here $l$ defines a characteristic material length, which accounts for size-dependency in this theory. Therefore, the relation Eq. (23) can be written as

$$\mu_{ij} = Q\delta_{ij} + 4Gl^2\left(\omega_{j,i} + c\omega_{i,j}\right) \tag{25}$$

We notice that the positive-definite elastic energy condition requires

$$3\lambda + 2G > 0, \qquad G > 0, \qquad \eta > 0, \qquad -1 < c < 1 \tag{26a-d}$$

The final governing equations for linear isotropic elastic material in this theory in terms of displacements become

$$\left[\lambda + G\left(1 + l^2\nabla^2\right)\right]u_{k,ki} + G(1 - l^2\nabla^2)\nabla^2 u_i = 0 \tag{27}$$

Koiter [4] considered the twist of the cylindrical bar in the framework of this theory. For pure torsion, the non-zero components of the curvature tensor $k_{ij}$ in cylindrical coordinates are

$$k_{rr} = -\frac{1}{2}\alpha \ , \quad k_{\theta\theta} = -\frac{1}{2}\alpha \ , \qquad k_{zz} = \alpha \tag{28}$$

If we assume $Q = 0$, the normal (torsional) couple-stress components in cylindrical coordinates become

$$\mu_{rr} = -2(\eta + \eta')\alpha = -2Gl^2(1+c)\alpha \tag{29a}$$

$$\mu_{\theta\theta} = -2(\eta + \eta')\alpha = -2Gl^2(1+c)\alpha \tag{29b}$$

$$\mu_{zz} = 4(\eta + \eta')\alpha = 4Gl^2(1+c)\alpha \tag{29c}$$

In this theory, the components of non-zero force-stresses are the same as those in classical theory, where

$$\sigma_{z\theta} = \sigma_{\theta z} = G\alpha r \tag{30}$$

We notice that the pure torsion does not satisfy the free surface boundary condition on the cylindrical surface, were the couple-stress $\mu_{rr} = -2Gl^2(1+c)\alpha$ exists. The exact solution has to be obtained apparently by solving the governing equations satisfying all boundary conditions. However, this will be impossible, because the indeterminate MTK-CST cannot satisfy the free surface boundary conditions in a systematic way. In fact, mathematical consistency requires that the couple stresses vanish. This means $\eta = \eta' = 0$, which reduces the couple stress theory to the classical theory of elasticity. Therefore, using this theory requires ignoring many inconsistencies [7]. This is the reason that the MTK-CST theory has been specified as indeterminate. However, we take this path by using the solution as an approximate elementary solution and investigate the overall effect of couple-stresses on torsional stiffness.

One might suggest that the total resultant torque at the ends of the bar as

$$T_\sigma + T_\mu \tag{31}$$

where

$$T_\sigma = \int_0^a \sigma_{z\theta} 2\pi r^2 dr = \frac{1}{2}\pi G\alpha a^4 \tag{32}$$

$$T_{\mu} = \int_{0}^{a} \mu_{zz} 2\pi r dr = \mu_{zz} \pi a^2 = 4\pi G\alpha (1+c) l^2 a^2 \tag{33}$$

We notice that $T_{\sigma}$ is the resultant torque in classical Cauchy elasticity. Therefore

$$T_{\sigma} + T_{\mu} = \frac{1}{2}\pi G\alpha a^4 + 4\pi G\alpha (1+c) l^2 a^2 \tag{34}$$

This can be written as

$$T_{\sigma} + T_{\mu} = \frac{1}{2}\pi G\alpha a^4 \left[1 + 8(1+c)\frac{l^2}{a^2}\right] \tag{35}$$

However, we must also consider the effect of the distribution of the normal couple-traction $\mu_{rr} = -2Gl^2 (1+c)\alpha$ on the surface of the cylinder by using the transformation method proposed by Koiter [4]. This transformation indicates that the distribution of the twisting surface couple-traction can be transformed into a line force traction $Gl^2 (1+c)\alpha$ tangent to the circular end edges. These also create an extra torsional moment $2\pi G\alpha (1+c) l^2 a^2$. This requires that the total torsional moment becomes

$$T = T_{\sigma} + T_{\mu} + 2\pi G l^2 a^2 (1+c)\alpha = \frac{1}{2}\pi G\alpha a^4 \left[1 + 12(1+c)\frac{l^2}{a^2}\right] \tag{36}$$

Therefore, for total torque, we obtain the expression

$$T = T_{\sigma}\left[1 + 12(1+c)\frac{l^2}{a^2}\right] \tag{37}$$

It should be noticed that this result can also be obtained by using the virtual work principle. Therefore, the torsional rigidity of the bar in this theory increases compared to classical theory, where the ratio

$$S = \frac{T}{T_{\sigma}} = 1 + 12(1+c)\frac{l^2}{a^2} \tag{38}$$

represents the normalized torsional rigidity. Since $-1 < c < 1$, we notice

$$1 < S < 1 + 24\frac{l^2}{a^2} \tag{39}$$

This result shows that in the original MTK-CST, there is a size-effect in pure twist of the circular bar.

### *3.3. Indeterminate symmetric modified couple stress theory (M-CST)*

In this theory originally proposed by Yang et al. [5], the couple stress tensor is symmetric, that is

$$\mu_{ji} = \mu_{ij} \tag{40}$$

and the curvature tensor is given by the symmetric tensor

$$\chi_{ij} = \frac{1}{2}\left(\omega_{i,j} + \omega_{j,i}\right) \tag{41}$$

which represents anticlastic deformation. In principal axes, these become pure torsional deformations.

The constitutive relations for linear isotropic elastic material are

$$\sigma_{(ij)} = \lambda e_{kk}\delta_{ij} + 2Ge_{ij} \tag{42}$$

$$\begin{aligned} \mu_{ij} &= Q\delta_{ij} + 8\eta\chi_{ij} \\ &= Q\delta_{ij} + 4\eta\left(\omega_{i,j} + \omega_{j,i}\right) \end{aligned} \tag{43}$$

We notice that the spherical part of the couple-stress tensor is still indeterminate, where $Q$ is a pseudo-scalar. The constant $\eta$ is the couple stress material constant accounting for size-dependency in this theory, where

$$\frac{\eta}{G} = l^2 \tag{44}$$

Therefore, the relation Eq. (43) can be written as

$$\mu_{ij} = Q\delta_{ij} + 4Gl^2\left(\omega_{j,i} + \omega_{i,j}\right) \tag{45}$$

Accordingly, the final governing equations for linear isotropic elastic material in this theory in terms of displacements become

$$\left[\lambda + G\left(1 + l^2\nabla^2\right)\right]u_{k,ki} + G(1 - l^2\nabla^2)\nabla^2 u_i = 0 \tag{46}$$

It should be noticed that in presenting the modified couple stress theory (M-CST), we have used different constitutive couple-stress coefficients to have more similarity to the other couple stress theories. The relations in the modified couple stress theory (M-CST) in its original form [5] can be found by scaling

$$\eta \to \eta/4 \quad (47a)$$

$$l \to l/2 \quad (47b)$$

Yang et al. [5] considered the twist of the cylindrical bar in the framework of this theory. For the pure torsion, the non-zero components of the symmetry tensor $\chi_{ij}$ in cylindrical polar coordinates are

$$\chi_{rr} = -\frac{1}{2}\alpha \ , \quad \chi_{\theta\theta} = -\frac{1}{2}\alpha \, , \qquad \chi_{zz} = \alpha \quad (48a\text{-}c)$$

If we assume $Q = 0$, the normal (torsional) couple-stress components in cylindrical coordinates become

$$\mu_{rr} = -4Gl^2\alpha \quad (49a)$$

$$\mu_{\theta\theta} = -4Gl^2\alpha \quad (49b)$$

$$\mu_{zz} = 8Gl^2\alpha \quad (49c)$$

The components of non-zero force-stresses are the same as those in classical theory, where

$$\sigma_{z\theta} = \sigma_{\theta z} = G\alpha r \quad (50)$$

We notice that as in the MTK-CST, the pure torsion also does not satisfy the free surface boundary condition on the cylindrical surface, were the couple-stress component $\mu_{rr} = -4Gl^2\alpha$ exists. Therefore, this pure torsion formulation is at most an approximate solution to this problem. The exact solution has to be obtained apparently by solving the governing equations satisfying all boundary conditions. However, this will be impossible, because the M-CST cannot satisfy the free surface boundary conditions in a systematic way. Mathematical consistency requires the couple-stresses to vanish, that is, $\eta = 0$. As a result, this theory reduces to the classical Cauchy elasticity.

However, for this solution as an approximation, we use the Koiter transformation method and obtain the effect of couple-stresses on torsional stiffness.

For this case, we notice

$$T_{\sigma} = \int_{0}^{a} \sigma_{z\theta} 2\pi r^{2} dr = \frac{1}{2}\pi G\alpha a^{4} \tag{51}$$

$$T_{\mu} = \int_{0}^{a} \mu_{zz} 2\pi r dr = 8\pi G\alpha l^{2} a^{2} \tag{52}$$

Therefore, the resultant torque at the ends of the bar seems to be

$$T_{\sigma} + T_{\mu} = \frac{1}{2}\pi G\alpha a^{4}\left(1+16\frac{l^{2}}{a^{2}}\right) \tag{53}$$

As before, we must add the effect of the normal couple-traction $\mu_{rr} = -4Gl^{2}\alpha$ on the surface of the cylinder by using the Koiter transformation method. This gives the extra torsional moment $4\pi G\alpha l^{2} a^{2}$. Therefore, the resultant torque becomes

$$T = T_{\sigma} + T_{\mu} + 4\pi G l^{2} a^{2}\alpha = \frac{1}{2}\pi G\alpha a^{4}\left(1+24\frac{l^{2}}{a^{2}}\right) \tag{54}$$

which can be expressed as

$$T = T_{\sigma}\left(1+24\frac{l^{2}}{a^{2}}\right) \tag{55}$$

This result can also be obtained directly by using the virtual work principle [5]. The relation Eq. (55) shows that the torsional rigidity of the bar in this theory increases compared to classical theory, such that the normalized torsional rigidity is

$$S = \frac{T}{T_{\sigma}} = 1 + 24\frac{l^{2}}{a^{2}} \tag{56}$$

### *3.4. Determinate skew-symmetric consistent couple stress theory (C-CST)*

In this theory [1], the couple stress tensor is skew-symmetric, that is

$$\mu_{ji} = -\mu_{ij} \tag{57}$$

and the curvature tensor is the skew-symmetric mean curvature tensor

$$\kappa_{ij} = \frac{1}{2}\left(\omega_{i,j} - \omega_{j,i}\right) \tag{58}$$

The constitutive relations for a linear isotropic elastic material are

$$\sigma_{(ij)} = \lambda e_{kk}\delta_{ij} + 2Ge_{ij} \tag{59}$$

$$\begin{aligned} \mu_{ij} &= -8\eta\kappa_{ij} \\ &= -4\eta\left(\omega_{i,j} - \omega_{j,i}\right) \end{aligned} \tag{60}$$

where there is no indeterminacy. The constant $\eta$ is the couple stress material constant accounting for size-dependency in this theory, where

$$\frac{\eta}{G} = l^2 \tag{61}$$

Therefore, the relation Eq. (60) can be written as

$$\mu_{ij} = -4Gl^2\left(\omega_{j,i} - \omega_{i,j}\right) \tag{62}$$

Consequently, the final governing equations for linear isotropic elastic materials in this theory in terms of displacements become

$$\left[\lambda + G\left(1 + l^2\nabla^2\right)\right]u_{k,ki} + G(1 - l^2\nabla^2)\nabla^2 u_i = 0 \tag{63}$$

We notice that for the pure twist of the cylindrical bar, the mean curvature tensor vanishes. Thus,

$$\kappa_{ij} = 0 \tag{64}$$

Since there is no mean curvature associated with pure twist deformation, no couple-stress is generated in the bar based on the consistent couple stress theory (C-CST), that is

$$\mu_{ij} = 0 \tag{65}$$

and the components of non-zero force-stresses are

$$\sigma_{z\theta} = \sigma_{\theta z} = G\alpha r \tag{66}$$

Therefore, the solution in C-CST reduces to that in classical theory, where the elastic pure twist of the cylindrical bar does not show any size effect. This character was first recognized in Reference [1]. We should emphasize that this solution is the exact solution to the governing Eq. (63), which satisfies all boundary conditions.

In this case, the resultant torque at the ends of the bar is exactly the same as the classical value. Thus,

$$T = T_{\sigma} = \int_0^a \sigma_{z\theta} 2\pi r^2 dr = \frac{1}{2}\pi G\alpha a^4 \tag{67}$$

We should emphasize that this interesting size-independency of pure torsion in the consistent skew-symmetric couple stress theory is not true for general torsion, where the deformation is not a pure twist. In particular, size effects can be seen in anisotropic, large deformation, non-quasi-static and dynamical conditions, where the torsional deformation is not pure and instead is accompanied by mean curvatures. For example, in elastodynamics, the torsional waves in the circular bar are size-dependent.

### *3.5. Summary*

Most interestingly, in the review of different couple stress theories for linear isotropic elasticity, we notice that the final governing Eqs. (27), (46) and (63) are the same. This is despite the fact that the constitutive relations in these theories are not the same and the response in pure torsion is quite different.

The resultant or total torque at the ends of the bar in different linear elastic theories can be summarized as

$$T = T_{\sigma}\left[1 + 12\left(1 + c\right)\frac{l^2}{a^2}\right] \tag{68}$$

where

$$-1 < c < 1 \qquad \text{for indeterminate MTK couple stress theory} \tag{69a}$$

$$c = 1 \qquad \text{for modified couple stress theory} \tag{69b}$$

$$c = -1 \qquad \text{for classical and consistent couple stress theories} \tag{69c}$$

Although these specifications make the comparison of results easy in this investigation, it does not mean that modified and consistent couple stress theories might be taken as special cases of the original indeterminate MTK-CST. This is obvious by noticing that these cases are excluded by the condition in Eq. (26) or Eq. (69) for the indeterminate MTK-CST. In addition, this similarity is only valid for isotropic materials. There is no simple analogy for general anisotropic cases. We should also emphasize that the curvature tensors are different from scratch in these theories.

To allow easy comparison of the deformation behavior of the circular bar in the different theories, we define the normalized torque $\tau$ and rotation $\gamma$ as follows:

$$\tau = \frac{2}{\pi}\frac{T}{a^3} \tag{70}$$

$$\gamma = \alpha a = \frac{a}{L}\phi \tag{71}$$

Therefore, the relation between normalized elastic torque and rotation in the different linear elastic theories can be expressed as

$$\tau = G\left[1 + 12\left(1+c\right)\frac{l^2}{a^2}\right]\gamma \tag{72}$$

Interestingly, we notice that the normalized rotation is the same as the engineering strain on the surface of the circular bar, where

$$\gamma = 2e_{z\theta} = \alpha a = \frac{a}{L}\phi \tag{73}$$

However, the normalized torque $\tau$ is the same as the shear stress on the surface of the bar only in the classical theory of elasticity and consistent skew-symmetric couple stress linear elasticity, corresponding to $c=-1$.

We also notice that the normalized elastic torsional rigidity $S$ can be generally represented as

$$S=\frac{\tau}{G\gamma}=1+12\left(1+c\right)\frac{l^2}{a^2} \tag{74}$$

Interestingly, for classical and consistent couple stress theories, where $c=-1$, we have

$$S=\frac{\tau}{G\gamma}=1 \tag{75}$$

and there is no size effect in pure torsion.

## 4. Comparison of torsion experiments with couple stress theories

Now we review details of the experimental results from previously published torsion tests on thin micro-diameter copper wires and compare these with the prediction from the three different couple stress theories.

Fleck et al. [9] developed a special screw driven system to conduct quasi-static torsion experiments on polycrystalline copper wires with sizes ranging from 12 to 170 μm in diameter. They employed a 60-mm-long glass filament to act as a torsional load cell. However, they presented their test results for the normalized torque-rotation in the plastic strain range, which comprised mostly of large rotation and large strains. Their results show a significant increase in the normalized torsional rigidity even in the small plastic strain range for the wires with five different diameters (12, 15, 20, 30, 170 μm). Although Fleck et al. [9] did not present the details of normalized torque-rotation relation in elastic range, from their plots, it seems there is no significant size effect for these copper wires in the elastic range.

Lu and Song [10] improved upon the experiment of Fleck et al. [9] by using a laser rotation sensor to measure the rotation angle of the glass fiber torque cell. Then, by using this technique, they characterized the size effect on torsional properties of copper wires with diameters ranging from 16 to 180 μm. The copper wires under investigation did not exhibit a significant size effect for four different diameters (16, 20, 30, 180 μm) in both elastic and plastic ranges. In these torsional experiments, all polycrystalline copper wires exhibit similar linear elastic response, followed by a non-linear plastic yielding at small strains. Significant oscillations are present in the plastic regime. However, the elastic shear modulus was calculated to be $G = 46.5\ \text{GPa}$, independent of the wire size. This test method was not capable of characterizing wires with diameter smaller than 15 μm due to insufficient sensitivity or resolution in torque measurement.

Liu et al. [8] also followed the Fleck et al. [9] concept with some modification in the technique and characterized the size effects on torsional properties of micro-diameter copper wires with diameters ranging from 20 to 50 μm. The normalized torque-rotation relationship from their experiments did not show significant size-effect for the wires with three different diameters (20, 25, 50 μm) in the elastic range. However, these experiments showed significant size effect in both the initial yielding and the plastic flow. Also, the response in the plastic range exhibited large oscillations.

In a new study, Song and Lu [11] improved their experimental technique further by a novel method developed by Walter and Kraft [12] to measure small torque for quasi-static torsional tests of copper wires with the diameter ranging from 12 to 30 μm. This technique provides more precise and straight-forward small torque measurement with sufficient resolution. Interestingly, the normalized torque-rotation results from their experiments showed again insignificant size effect for the wires with four different diameters (12, 16, 20, 30 μm) in both the elastic and plastic ranges. However, the stresses after yield are approximately 20% lower than those presented previously by Lu and Song [10]. We notice that this may be because the shear strain rate in this study is more precisely controlled, which results in more accuracy in the torque measurement, than in their previous experiments. Despite this improved control, significant oscillations are still present in the plastic regime.

We conclude that experiments by Song and Lu [11], Liu et al. [8] and Lu and Song [10] show no size effect in the elastic range for pure torsion of micro-diameter copper wires in quasi-static loading. This agrees completely with the prediction of the consistent skew-symmetric couple stress theory. Interestingly, we notice that the consistent skew-symmetric couple stress theory is the only size-dependent theory, which predicts no size effect for linear elastic wires in pure torsional deformation under quasi-static conditions. Therefore, this theory not only is mathematically self-consistent, but also is consistent with the results of physical torsion experiments in the linear elastic range.

One might suggest that it would be good for the sake of comparison to evaluate the anticipated size effect on the torsional rigidity in the other size-dependent theories, such as MTK-CST and M-CST. Interestingly, Bansal et al. [13] have estimated the length scale parameter $l$ in consistent couple stress theory (C-CST) for four different single crystals (NaCl, KCl, Cu, CuZn) based on available wave velocities from ultrasonic excitation and phonon dispersion curves, along with adiabatic bulk moduli measurements. For face-centered cubic (fcc) single crystal copper (Cu), the length scale parameter has been estimated to be $l = 23\ \mu m$. One may expect that this length scale parameter would be on the same order for a polycrystalline copper, which can be considered isotropic. Therefore, we may use $l = 23 \mu m$ for copper as an estimate of the length scale parameter in consistent couple stress theory. For the shear modulus, we use $G = 46.5$ GPa from Lu and Song [10], while for Poisson ratio, we may use the well-established value $\nu = 0.34$. Since all couple stress theories (original MTK-CST, symmetric M-CST and skew-symmetric C-CST) give very similar results in ultrasonic excitation, phonon dispersion curves and bending problems, it is reasonable to use this length scale parameter value $l = 23\ \mu m$ for copper in all of these theories. However, in the original MTK-CST, we need the second couple stress parameter $\eta'$ or $c$. For the comparisons, here we take $c = 0$, which means $\eta' = 0$.

Table 1 summarizes the theoretical normalized elastic torsional rigidity $S$ calculated from Eq. (74) with $l = 23\ \mu m$ in the four different theories for copper wires with diameters ranging from 12 to 180 μm, as used in the aforementioned experiments. It is interesting to notice that the normalized torsional rigidity $S$ becomes much greater than one hundred for the smaller diameters in both the original MTK-CST and M-CST. Clearly, these results are in stark contrast to the experimental results, which show no size effect at all.

However, we must recognize that this characteristic length scale is not as well-established as the classical elastic parameters $E$ and $\nu$. Consequently, let us explore the consequences of dramatically reduced values of this length scale. To compensate for the uncertainty in the value of the length scale parameter from Bansal et al. [13], we also calculate $S$ for lesser orders of magnitude for $l$. For example, Table 2 shows that the normalized elastic torsional rigidity $S$ for $l = 2.3 \mu m$, which is one tenth of the original length scale parameter value. Notice that the torsional rigidity $S$ still would become greater than twice the classical result for the smaller diameters. Meanwhile, Table 3 provides the corresponding values for $S$ with $l = 0.23 \mu m$. If $l$ for copper is at this lower level examined in Table 3, then the size effects in torsion would be difficult to detect in experiments with diameters on the order of tens of microns.

**Table 1** Normalized elastic torsional rigidity $S$ for copper with $G = 46.5\text{GPa}$, $\nu = 0.34$, $l = 23\ \mu m$.

| Wire diameter $d = 2a$ | Experiments | Classical elasticity $c = -1$ | MTK couple stress elasticity for $c = 0$ | Modified couple stress elasticity $c = 1$ | Skew-symmetric couple stress elasticity $c = -1$ |
|---|---|---|---|---|---|
| 180 μm | 1 | 1 | 1.784 | 2.567 | 1 |
| 170 μm | 1 | 1 | 1.879 | 2.757 | 1 |
| 30 μm | 1 | 1 | 29.21 | 57.43 | 1 |
| 25 μm | 1 | 1 | 41.63 | 82.25 | 1 |
| 20 μm | 1 | 1 | 64.48 | 127.96 | 1 |
| 16 μm | 1 | 1 | 100.2 | 199.4 | 1 |
| 15 μm | 1 | 1 | 113.9 | 226.7 | 1 |
| 12 μm | 1 | 1 | 177.3 | 353.7 | 1 |

**Table 2** Normalized elastic torsional rigidity $S$ for copper with $G = 46.5\text{GPa}$ , $\nu = 0.34$ , $l = 2.3\ \mu m$ .

| Wire diameter $d = 2a$ | Experiments | Classical elasticity $c = -1$ | MTK couple stress elasticity $c = 0$ | Modified couple stress elasticity $c = 1$ | Skew-symmetric couple stress elasticity $c = -1$ |
|---|---|---|---|---|---|
| 180 µm | 1 | 1 | 1.008 | 1.016 | 1 |
| 170 µm | 1 | 1 | 1.009 | 1.018 | 1 |
| 30 µm | 1 | 1 | 1.282 | 1.564 | 1 |
| 25 µm | 1 | 1 | 1.406 | 1.813 | 1 |
| 20 µm | 1 | 1 | 1.635 | 2.270 | 1 |
| 16 µm | 1 | 1 | 1.992 | 2.984 | 1 |
| 15 µm | 1 | 1 | 2.129 | 3.257 | 1 |
| 12 µm | 1 | 1 | 2.763 | 4.527 | 1 |

**Table 3** Normalized elastic torsional rigidity $S$ for copper with $G = 46.5\text{GPa}$ , $\nu = 0.34$ , $l = 0.23\mu m$ .

| Wire diameter $d = 2a$ | Experiments | Classical elasticity $c = -1$ | MTK couple stress elasticity $c = 0$ | Modified couple stress elasticity $c = 1$ | Skew-symmetric couple stress elasticity $c = -1$ |
|---|---|---|---|---|---|
| 180 µm | 1 | 1 | 1.000 | 1.000 | 1 |
| 170 µm | 1 | 1 | 1.000 | 1.000 | 1 |
| 30 µm | 1 | 1 | 1.003 | 1.006 | 1 |
| 25 µm | 1 | 1 | 1.004 | 1.008 | 1 |
| 20 µm | 1 | 1 | 1.006 | 1.013 | 1 |
| 16 µm | 1 | 1 | 1.010 | 1.020 | 1 |
| 15 µm | 1 | 1 | 1.011 | 1.023 | 1 |
| 12 µm | 1 | 1 | 1.018 | 1.035 | 1 |

We should notice that in skew-symmetric consistent couple stress theory (C-CST), the twist of a non-linear isotropic elastic circular bar is also a pure torsional deformation in static or quasi-static loading. Therefore, there also should not be any size effect for this case in a physical experiment. Although a plasticity formulation based on the C-CST has not been developed yet, we expect that this plasticity theory with an isotropic hardening (now strain and mean curvature hardening) model should not predict any size effect for pure torsion of micro-diameter wires in the plastic range. This is because there is no mean curvature in a quasi-static condition, as long as small deformation theory is valid. It seems this can explain the experimental results from Song and Lu [10, 11] that there is no size effect, even in the non-linear plastic yielding of micro-diameter copper wires.

However, when the hardening is anisotropic, there is no pure torsion in the bar in the plastic range, which may exhibit some size effect. We would also suggest that plastic deformation is more sensitive to the end conditions and the quasi-static loading. This means that it is practically more difficult to have a pure quasi-static twist in the bar in the plastic range, even with isotropic hardening. The significant oscillations apparent in the plastic regime in all of the recent pure torsion experiments provide strong support for this inherent sensitivity. We should also mention that the hardening of material in small sizes might become anisotropic, as a result of the propagation of end effects, which might trigger an inhomogeneous distribution of dislocations and disclinations. Therefore, these effects might cause deviation from a pure torsional deformation, which in turn could result in the appearance of a size effect. These hypothesized causes might explain the size effect phenomenon in the torsion of the cylindrical bar in the plastic range, which was observed experimentally by Fleck et al. [9] and Liu et al. [8]. Consequently, skew-symmetric consistent couple stress theory (C-CST) may also be capable of explaining size effect phenomena in the torsion of the circular cylindrical bar in the plastic range. However, this size-dependent couple stress plasticity remains to be developed.

## 5. Conclusion

Recent experiments by Song and Lu [11], Liu et al. [8] and Lu and Song [10] to characterize torsion of micro-diameter copper wires in quasi-static loading show no significant size effect in the elastic range. These results align well with the prediction of skew-symmetric consistent couple stress theory (C-CST) that there is no size effect for torsion of an isotropic elastic circular bar in quasi-static loading. This becomes more important when we notice that skew-symmetric couple stress theory is the only size-dependent theory that predicts size independent torsional response. Therefore, this theory not only is mathematically consistent, but is in line with recent torsion experiments on micro-diameter wires in the elastic range.

These experiments also tend to refute the results from other size-dependent theories, such as indeterminate MTK-CST and M-CST, from an experimental standpoint. First of all, simple pure torsion is not an exact solution in these size-dependent theories. This is because in these theories the pure torsion deformation creates a normal twisting couple-traction on the free surfaces [7]. The true solution must be obtained by solving the governing boundary value problems satisfying all boundary conditions. However, this is not possible because these theories are inconsistent from a fundamental perspective. Mathematical consistency in the torsion problem requires that these theories reduce to classical Cauchy elasticity. On the other hand, if we use the pure solution as an approximate solution in these theories, unrealistically significant size effect in torsion of micro-diameter copper wires may be predicted, depending upon the length scale. We have demonstrated this for MTK and modified couple stress theories by using a broad range of values for the length scale parameter $l$ for isotropic polycrystalline micro-diameter copper wires. These theories show the normalized elastic torsional rigidity $S$ can become greater than one hundred for smaller diameters, which clearly contradicts the experimental data. Only at very small characteristic lengths on the order of hundreds of nanometers do the MTK-CST and M-CST exhibit negligible size-effects in torsion.

Although a plasticity theory based on the consistent couple stress theory has not been developed yet, we expect that the corresponding size-dependent plasticity theory with an isotropic hardening should also predict no size effect for pure torsion of micro-diameter wires in quasi-static plastic range. This can explain the experimental results from Song and Lu [10, 11] that there is no size effect in the non-linear plastic yielding of micro-diameter copper wires. However, any deviation from pure torsional deformation, such as sensitivity to the end constraints and loading conditions, anisotropic hardening and elastodynamic effects, could result in the appearance of size effects. In particular, we would suggest that the torsion experiment for the bar in the plastic range is more sensitive to these factors compared to the response in the elastic range. The noticeable oscillations in the plastic regime in all of the recent experiments tend to support this conjecture.

Finally, we should emphasize the need for carefully controlled bending experiments to establish reliable values for the couple stress characteristic material length parameters for polycrystalline copper, as well as for other materials. If the length scales are found to be sufficiently large (e.g., $l$ equal to several microns or larger for copper), then this would confirm the predictions of consistent couple stress theory, while eliminating the inconsistent indeterminate couple stress theory (MTK-CST) and indeterminate modified couple stress theory (M-CST) as viable models to capture mechanical behavior at small continuum scales. If the length scales are below the micron level, then MTK-CST and M-CST would remain suspect nonetheless due to the theoretical shortcomings of these theories.

## References

1. Hadjesfandiari, A.R., Dargush, G.F. Couple stress theory for solids. Int. J. Solids Struct. **48**, 2496-2510 (2011)
2. Hadjesfandiari, A.R., Dargush, G.F. Foundations of consistent couple stress theory. Preprint arXiv: 1509.06299 (2015)
3. Mindlin, R.D., Tiersten, H.F. Effects of couple-stresses in linear elasticity. Arch. Rational Mech. Anal. **11**, 415–488 (1962)
4. Koiter, W.T. Couple stresses in the theory of elasticity, I and II. Proc. Ned. Akad. Wet. (B) **67**, 17-44 (1964)
5. Yang, F., Chong, A.C.M., Lam, D.C.C., Tong P. Couple stress based strain gradient theory for elasticity. Int. J. Solids Struct. **39**, 2731–2743 (2002)
6. Lazopoulos, K.A. On bending of strain gradient elastic micro-plates. Mech. Res. Commun. **36**, 777–783 (2009)
7. Hadjesfandiari, A.R., Dargush, G.F. Evolution of generalized couple-stress continuum theories: a critical analysis. Preprint arXiv: 1501.03112 (2015)
8. Liu, D., He, Y., Dunstan, D.J., Zhang, B., Gan, Z., Hu, P., Ding, H. Toward a further understanding of size effects in the torsion of thin metal wires: an experimental and theoretical assessment. Int. J. Plasticity **41**, 30-52 (2013)
9. Fleck. N.A., Muller, G.M., Ashby, M.F., Hutchinson, J.W. Strain gradient plasticity: theory and experiment. Acta Metall. Mater. **42**, 475–487 (1994)
10. Lu, W.-Y., Song, B. Quasi-static torsion characterization of micro-diameter copper wires. Exp. Mech. **51**, 729-737 (2011)
11. Song, B., Lu, W.-Y. An improved experimental technique to characterize micro-diameter copper wires in torsion. Exp. Mech. **55**, 999-1004 (2015)

12. Walter, M., Kraft, O. A new method to measure torsion moments on small-scaled specimens. Rev. Sci. Instrum. **82**, 035109 (2011)
13. Bansal, D., Dargush, G.F., Aref, A.J., Hadjesfandiari, A.R. Cubic single crystal representations in classical and size-dependent couple stress elasticity. Preprint arXiv:1510.06048 (2015)